\RequirePackage{lineno}
\documentclass[aps, prl, twocolumn,print]{revtex4}

\usepackage{graphicx}
\usepackage{amssymb,amsmath, wasysym}

\newcommand{\icm}[1]{\ensuremath{#1\ \mathrm{cm^{-1}}}}
\newcommand{\fs}[1]{\ensuremath{#1\ \mathrm{fs}}}
\newcommand{\ps}[1]{\ensuremath{#1\ \mathrm{ps}}}

\begin{document}
\title{Nonlinear frequency conversion is controlled by vacuum fluctuations}

\author{Emre Y\"uce,$^{1,2,\ast}$ Georgios Ctistis,$^{1,3}$ Julien Claudon,$^{4,5}$  Jean-Michel G\'erard,$^{4,5}$ Willem L. Vos$^{1}$}
\affiliation{
$^{1}$Complex Photonic Systems (COPS), MESA+ Institute for Nanotechnology, University of Twente, P.O. Box 217, 7500 AE Enschede, The Netherlands \\
$^{2}$ Programmable Photonics Group, Department of Physics, Middle East Technical University, 06800 Ankara, Turkey \\
$^{3}$NanoBioInterface Group, Department of Nanotechnology, School of Life Science, Engineering, and Design, Saxion University of Applied Sciences, 7500 KB Enschede, The Netherlands. \\
$^{4}$Univ. Grenoble Alpes, INAC-PHELIQS, 38000 Grenoble, France\\
$^{5}$CEA, INAC-PHELIQS, Nanophysics and Semiconductors Laboratory, 38000 Grenoble, France\\
}
\date{October $10^{th}$ 2016}

\maketitle

{\bfseries
Ever since the advent of nonlinear optics, the generation of light by frequency-conversion is drawing continued attention, and leading to emerging applications such as supercontinuum sources for ultra stable clocks and advanced microscopy. 
A modern approach to frequency-conversion is to switch light confined in micro- and nanocavity resonances to enable on-chip operation.
Supposedly, nonlinear frequency conversion in such confined media differs from traditional non-linear optics in three key features regarding output spectrum, frequency shift, and critical time scale. 
Therefore, we switch GaAs-AlAs microcavities by the electronic Kerr effect, and study a range of quality factors to bridge the confined and the traditional non-linear regimes. 
We uncover the key role of the density of vacuum fluctuations, \textit{i.e.}, the local density of optical states (LDOS) for newly generated frequencies, a concept inspired by cavity quantum electrodynamics (cQED). 
As a result, we succeed to establish a framework, which not only describes nonlinear optics both in traditional bulk and in confined media but also opens a new control dimension in changing the color of light.
}

\section{Introduction}\label{sec:introduction}

The generation of optical frequencies starting from incident light is well-known in nonlinear optics~\cite{stolen.1978.pra,dudley.2006.rmp, boyd.2008.book, kampfrath.2010.pra}. 
In self-phase modulation~\cite{stolen.1978.pra}, a laser beam induces a nonlinear polarization in a spatially homogeneous nonlinear medium. 
The polarization leads to a temporal modulation of the phase $\phi(t)$ of the incident wave that results in an instantaneous frequency shift $\delta \omega(t)$ of the output light equal to the rate of change of the phase $\delta \omega(t) = -\partial\delta \phi(t) / \partial t$. 
Since the frequency shift increases with incident light intensity and with interaction length, long fibers have become popular media to obtain frequency shifts by means of the electronic Kerr effect~\cite{stolen.1978.pra, boyd.2008.book}. 

In modern micro and nano-confined structures, however, the frequency of confined light is nonlinearly converted by quickly changing the cavity resonance in time in confined media~\cite{notomi.2006.pra, dong.2008.prl, tanabe.2009.prl, preble.2007.nat.photon, mccutcheon.2007.oe, eilenberger.2013.SciRep, Moille.2016.lpr, Chai.2016.AdvOptMat, Song.2018.AdvOpMat}.
As lucidly explained in Refs.~\cite{notomi.2006.pra, dong.2008.prl, tanabe.2009.prl}, the physics of frequency conversion in confined media supposedly differs from traditional nonlinear optics in at least three key features: 

\noindent
(1) The output spectrum reveals a discrete peak at each resonance ($\omega = \omega_{c}$) instead of a continuum as in traditional nonlinear optics.

\noindent
(2) The frequency shift $\delta \omega(t)$ depends on the tuning rate of the resonance $\omega_{c}(t)$, but not on the tuning rate of the phase ($\partial\delta \phi(t) / \partial t$), as in traditional nonlinear optics. 

\noindent
(3) The switch process has a critical time scale $\partial t$ that should be faster than the resonance storage time ($\partial t < \tau_{c}$), a situation that does not pertain to traditional nonlinear optics.

The central question that we address here is how to reconcile the apparent differences regarding frequency conversion in traditional homogeneous media with conversion in nanophotonic resonant cavities. 
Therefore, we have initiated a study of the dynamics of the frequency conversion of light confined in GaAs-AlAs microcavities, by pump-probe reflectivity~\cite{diels.2006.book}. 
The cavity resonance is switched in time by the electronic Kerr effect that represents the fastest possible, completely reversible, and purely dispersive all-optical switch~\cite{ctistis.2011.apl}. 
The key feature of our paper is to consider the local density of optical states (LDOS), a concept widely used in spontaneous emission control~\cite{sprik.1996.epl, novotny.2006.book}. 
By varying the quality factor $Q$ of the microcavities, we control both the LDOS on cavity resonance, and the continuum LDOS of the surrounding vacuum that leaks into the cavity. 

\section{Experimental techniques}\label{sec:expttechniques}

The setup for the ultrafast pump-probe experiments~\cite{euser.2009.rsi} is schematically illustrated in Figure~\ref{SetupSamp}(a) and described in detail in the Supplementary Material.
In brief, the pump pulses are tuned to a low frequency $\omega_{pu}= 4165 \ \rm{cm^{-1}}$ $(\lambda_{pu}=2400 \rm{nm})$ to optimize ultrafast electronic Kerr switching and avoid slow and absorptive free-carrier effects~\cite{hartsuiker.2008.jap,yuce.2016.oe}. 
The frequency of the probe light $\omega_{pr}$ is tuned to the cavity resonance frequency $\omega_{c}$. 
The reflected signal from the cavity is frequency-resolved with a spectrometer. 
The measured transient reflectivity is a result of the probe light that impinges at delay $\Delta t$ and circulates in the cavity during, on average, the storage time $\tau_{c}$. 
The relatively slow detector integrates the short pulses ($\tau_{P} = 140 \pm 10 \ \rm{fs}$). 
The measured transient reflectivity signal therefore contains information on the cavity resonance during the storage of the probe light in the cavity and it should thus not be confused with the instantaneous reflectivity at the delay $\Delta t$, see Ref.~\cite{yuce.2016.oe}. 

\begin{figure}[htb]
\begin{center}
\includegraphics[width=0.9\linewidth]{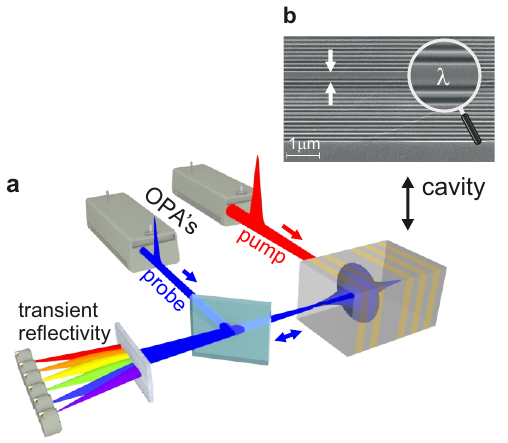}
\caption{\emph{
(a) Schematic of the switch setup. 
The pump and probe pulses emanate from two optical parametric amplifiers (OPAs). 
The probe beam path is shown in blue, the pump beam path in red. 
The detected transient reflectivity of the cavity is spectrally resolved. 
(b) Scanning electron micrograph (SEM) of the cross-section of a GaAs-AlAs microcavity. 
GaAs layers appear light grey, and AlAs layers dark grey. 
The magnifier highlights the $\lambda$-layer, with a thickness indicated by the white arrows.
The GaAs substrate appears at the bottom. 
}}
\label{SetupSamp}
\end{center}
\end{figure}

\begin{figure}[htb]
\begin{center}
\includegraphics[width=1.0\linewidth]{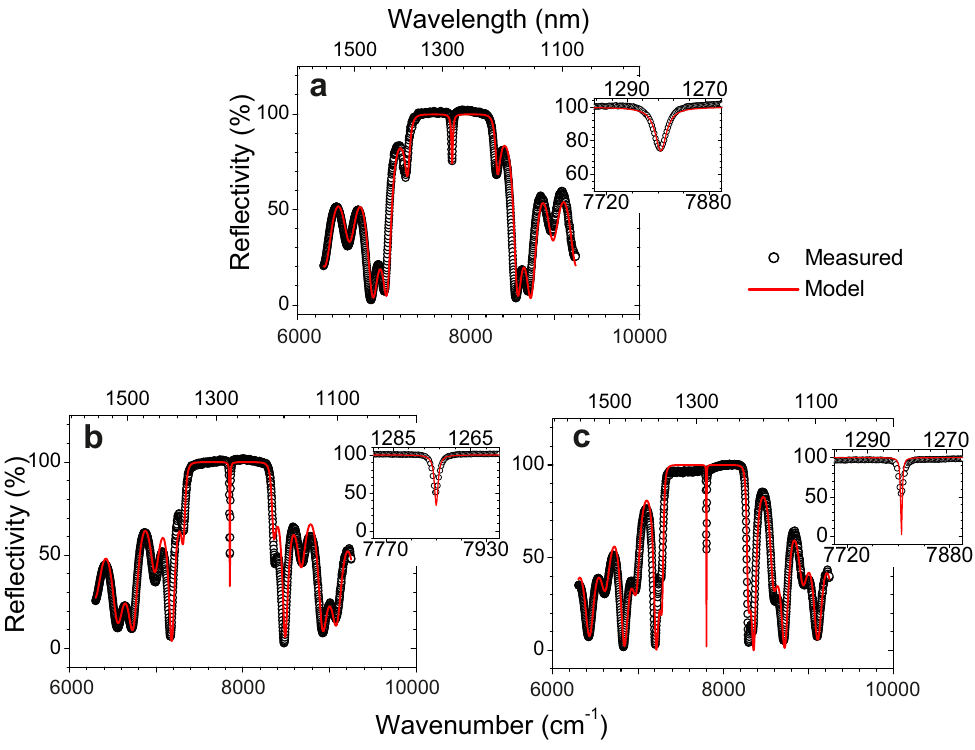}
\caption{\emph{Measured (black circles) and calculated (red curves) reflectivity spectra of the microcavities with increasing number of top layers: (a) 7 pairs with $Q = 390 \pm 60$,
(b) 9 pairs with $Q = 540 \pm 60$, and 
(c) 15 pairs with $Q = 890 \pm 60$. 
The ordinates show the reflectivity, the bottom abscissas the frequency and the top abscissas the wavelength. 
Within each stop band a narrow trough signals the cavity resonance $\omega_{0}$, shown with a higher resolution in the inset of each panel. 
The calculations are performed with the transfer matrix model. 
}}
\label{LinRef}
\end{center}
\end{figure}

We have performed experiments on planar microcavities that are made from GaAs and AlAs layers by molecular beam epitaxy, as shown in Figure~\ref{SetupSamp}(b).~\cite{cho.1975.PSSC, farrow.1995}. 
The microcavities consist of a GaAs $\lambda$-layer ($d=376 \ \rm{nm}$) sandwiched between two Bragg mirrors made of $\lambda/4$-thick layers of GaAs ($d_{GaAs}=94 \ \rm{nm}$) and AlAs ($d_{AlAs}= 110 \ \rm{nm}$), respectively. 
The bottom Bragg mirror of the cavities consists of 19 pairs of GaAs and AlAs and the top Bragg mirror has 15 pairs of layers. 
All structures were made from the same parent structure to minimize sample-to-sample variations.
In order to prepare several samples with reduced cavity storage times, the
parent sample is cut into smaller chips of $5 \ \textrm{mm} \times 5 \ \textrm{mm}$.
The edge of the chips is protected with optical resist. 
Subsequently, a number of layers is selectively removed from the top Bragg mirror by dry- and wet-etching techniques to obtain cavities with reduced quality factors. 
The samples are first etched by reactive ion-etching (RIE) that reacts with both GaAs and AlAs layers. 
The depth of the etching process is controlled interferometrically with a $20$ nm accuracy during the RIE step. 
After removing the targeted number of layers with RIE, the remaining AlAs layer on top of the structure is selectively removed by wet etching using HF.
We varied the number of pairs on the top Bragg mirror between 7, 9, and 15 and thus the reflectivity of the top Bragg mirror, respectively. As a result, we obtain cavities with quality factors ranging from $Q=890 \pm 60$, $Q=550 \pm 60$, to $Q=390 \pm 60$,  corresponding to storage times between \fs{605 \pm 45} and \fs{265 \pm 45}. 
Figure~\ref{LinRef} shows the reflectivity spectra of all cavities.  
The reflectivity at the cavity resonance is large given the asymmetric mirror reflectivities. 
As the cavity becomes more symmetric, the reflectivity at the cavity resonance trough decreases.
A greater thickness of the top Bragg mirror increasingly shields the cavity from the local density of optical states (LDOS) of the surrounding vacuum that tunnels into the microcavity~\cite{hasan.2018.prl}.

\section{Results: ultrafast all-optical switching}\label{sec:results}

\begin{figure}[htb]
\begin{center}
\includegraphics[width=0.95\linewidth]{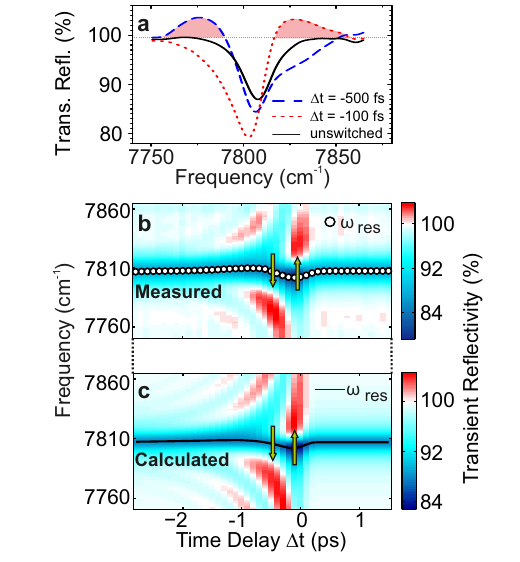}
\caption{\emph{
Observations of frequency conversion of light confined in a $Q = 390$ microcavity.
(a) Transient reflectivity spectra $R^t(\omega)$ measured at different time delays, and unswitched reference at $\Delta t=\ps{2.8}$ where pump and probe pulses do not overlap.
The unswitched cavity resonance is at $\omega_{c}/2\pi c=\icm{7807.5}$ (or $\lambda_{c} = 1280.8$ nm). 
The reflectivity minimum on resonance ($R^t_{min}=85\ \%$) is high due to the asymmetric cavity design~\cite{yuce.2016.oe}. 
Newly converted light is apparent as $R^t(\omega) > 100 \%$ (pink filling). 
(b) Map of the measured and (c) map of the calculated transient reflectivity versus time delay. 
Green arrows indicate the direction of frequency conversion. 
White symbols (b) and the solid curve (c) mark the dynamic resonance frequency $\omega_{c}(t)$ of the cavity. }}
\label{SingleSwitch}
\end{center}
\end{figure}

Figure~\ref{SingleSwitch}(a) presents spectra measured at several pump-probe delays $\Delta t$. 
During the switch event, the cavity resonance rapidly shifts by \icm{5.7} to a lower frequency at maximum pump-probe overlap, in agreement with the positive non-degenerate Kerr coefficient of GaAs~\cite{yuce.2012.josab}. 
The resonance returns to the starting frequency immediately after the pump pulse has gone. 
The overlap is maximal at $\Delta t=\fs{-100}$, since it takes incident probe light about $100 \ \rm{fs}$ to charge the cavity. 
Escaping frequency-converted light is collected in reflection geometry, where it interferes with probe light that directly reflects from the top Bragg mirror.
The interference corresponds to a phase-sensitive heterodyning that allows us to sensitively detect frequency-converted light outside the instantaneous resonance, and to observe signal over a much longer time (up to \ps{2}) than the cavity storage time ($\tau_{c} = \fs{300}$). 

The detected signal is the frequency-resolved transient reflectivity $R^t (\omega)$~\cite{euser.2009.rsi}. 
For the unswitched cavity $R^t (\omega) = 100 \%$ outside the resonance. 
Figure~\ref{SingleSwitch}(a) reveals a remarkable transient reflectivity $R^t (\omega) > 100 \%$ at $\Delta t=\fs{-100}$ delay and at frequencies \emph{above} the cavity resonance. 
Simultaneously and in keeping with energy conservation, the signal is depleted at frequencies below the cavity resonance, thus indicating a blue shift of confined light. 
At another delay $\Delta t=-\fs{500}$, excess transient reflectivity $(R^t (\omega) > 100 \%)$ occurs \emph{below} the cavity resonance with a simultaneous depletion above the resonance, thus signaling red-conversion of confined light. 
The cavity-confined light is either blue-converted or red-converted by as much as \icm{50}, much greater than the $\delta \omega_{c} = \icm{5.7}$ shift of the cavity resonance. 
This spectral observation contradicts the previous nanophotonic view of the output being peaked at the cavity resonance, as per key feature $\# 1$.

Figure~\ref{SingleSwitch}(b) shows a map of the measured transient reflectivity spectra $R^t (\omega)$ versus pump-probe delay $\Delta t$. 
Blue-converted light occurs at time delays between \fs{-100} and \fs{+100}.
Here the cavity resonance frequency $\omega_{c}$ increases, thus the confined light experiences a decreasing refractive index, and a concomitant negative phase change $(\partial\delta \phi(t) / \partial t < 0)$. 
Red-shifted light appears between $\Delta t = \fs{-900}$ and \fs{-100}, where $\omega_{c}$ decreases.
Here the confined light experiences an increasing refractive index, and thus a positive phase change $(\partial\delta \phi(t) / \partial t > 0)$.
Near $\Delta t < \fs{-500}$, both blue- and red-shifted light appear, since the refractive index both increases and decreases while the probe light circulates in the cavity.
Here, the magnitude of the red-shifted light is larger than for the blue-shifted light, since the pump pulse first induces an increasing and then a decreasing refractive index while the stored light is escaping from the cavity. 

To confirm these observations, we have calculated the nonlinear frequency conversion shown in Figure~\ref{SingleSwitch}(c) using a time-resolved model that has no free parameters~\cite{harding.2012.josab, yuce.2016.oe} (see Supplementary Material, Theory.)
Our calculated spectra reveal blue- and red-converted light at the same frequencies, same bandwidth, and same timing as in the measurements shown in Figure~\ref{SingleSwitch}(b). 
As the timing of the blue- or red-conversion corresponds to the occurrence of a positive or a negative rate of change of the phase $(\partial\delta \phi(t) / \partial t)$, see key feature $\# 2$, one might naively conclude that the cause of the frequency shift is settled in favor of traditional phase modulation, even though we employ a microcavity. 
In the next section, we will see that this conclusion is invalid.

\section{Role of the local density of states}\label{sec:ldos}

\begin{figure}[htb]
\begin{center}
\includegraphics[width=0.850\linewidth]{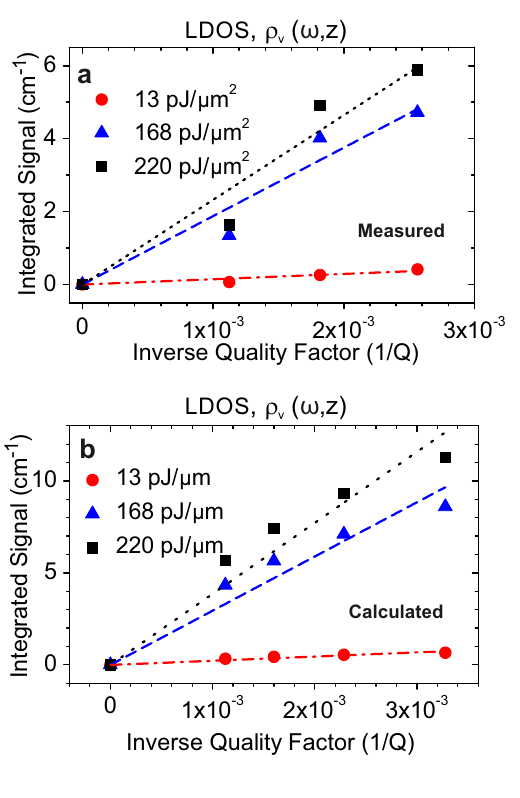}
\caption{\emph{(a) Measured and (b)calculated integrated blue-converted light, averaged over time delay between $\Delta t=-100$ and $\fs{0}$ and integrated over frequency, versus inverse quality factor $1/Q$.
The top abscissa gives the corresponding local density of states of the vacuum $\rho_v(\omega,z_c)$ that tunnels into the cavity through the Bragg mirrors. 
Data are shown for pump pulse fluences 13, 168, and 220 $pJ / \mu m^2$ (circles, triangles, squares), and lines are linear fits to the data, respectively. 
}}
\label{DOSMeasCalc}
\end{center}
\end{figure}

Figures~\ref{DOSMeasCalc} (a) and (b) show the integrated measured and calculated blue-converted signal as a function of inverse quality factor $(1/Q)$, respectively. 
We observe an excellent agreement between the measured and calculated intensities of frequency converted light.
It is remarkable that the amount of frequency-converted light increases as much as $3 \times$ with increasing inverse $Q$-factor in both our measurements and calculations. 
At this point one might hypothesize that the dramatic increase is due to the larger (broadband frequency integrated) internal intensity in a lower $Q$-factor cavity. 
However, we verified that the change of integrated internal intensity is only $5\%$ within the range of $Q$-factors studied here (see Supplementary Material), since the broader bandwidth in a low-$Q$ cavity is offset by the lower maximum field amplitude. 
Hence, this hypothesis is invalid. 

In nanophotonics, the quality factor $Q$ is known to gauge the local density of optical states (LDOS) \textit{on resonance} at the cavity center $\rho_c(\omega_{c},z_c)$, that plays a central role in the Purcell effect in cQED~\cite{purcell.1946.pr, gerard.1998.prl}. 
Thus, the trend in Figures~\ref{DOSMeasCalc} (a, b) would imply that the frequency-converted signal decreases with increasing LDOS, which is opposite to common lore in nanophotonics where a signal typically increases with increasing LDOS. 
The solution to this paradox is rooted in our realization that the LDOS \emph{at frequencies outside the resonance} can be identified with $1/Q$~(see Supplementary Material, Theory). 
This LDOS is the density of the vacuum fluctuations $\rho_v(\omega,z_c)$ that surround the cavity structure and that tunnel into the cavity through the Bragg mirrors~\cite{hasan.2018.prl}. 
Therefore, our observations in Figure~\ref{DOSMeasCalc} imply that the color-converted intensity increases with the vacuum LDOS that leaks into the central layer where light is generated, mostly at frequencies outside the resonance (see Fig.~\ref{SingleSwitch}). 
The increasing intensity of frequency-converted light with $1/Q$ does not agree with the traditional nonlinear argument (namely rate of change of the phase), since the rate of change is the same for all cavities, regardless of $Q$. 
Our identification that the LDOS plays a central role in nonlinear color-conversion introduces a novel concept in nonlinear optics.

Fig.~\ref{DOSMeasCalc} also shows that the magnitude of frequency-converted light increases with pump pulse fluence. 
This is understood from the third-order electronic-Kerr nonlinearity. 
The induced nonlinear polarization is expanded in the nonlinear index $n = n_{0} + n_{2}\cdot I_{pu}$ that increases linearly with pump fluence $I_{pu}$.\cite{boyd.2008.book} 
At constant pump pulse duration, an increased refractive index corresponds to an increased rate of change of the phase $(\partial\delta \phi(t) / \partial t)$, and thus an increased frequency shift. 
Regarding key feature $\# 2$, we thus conclude that the frequency shift and intensity are related to \textit{both} the LDOS and the rate of change of the phase. 

\begin{figure*}[htb]
\begin{center}
\includegraphics[width=0.8\linewidth]{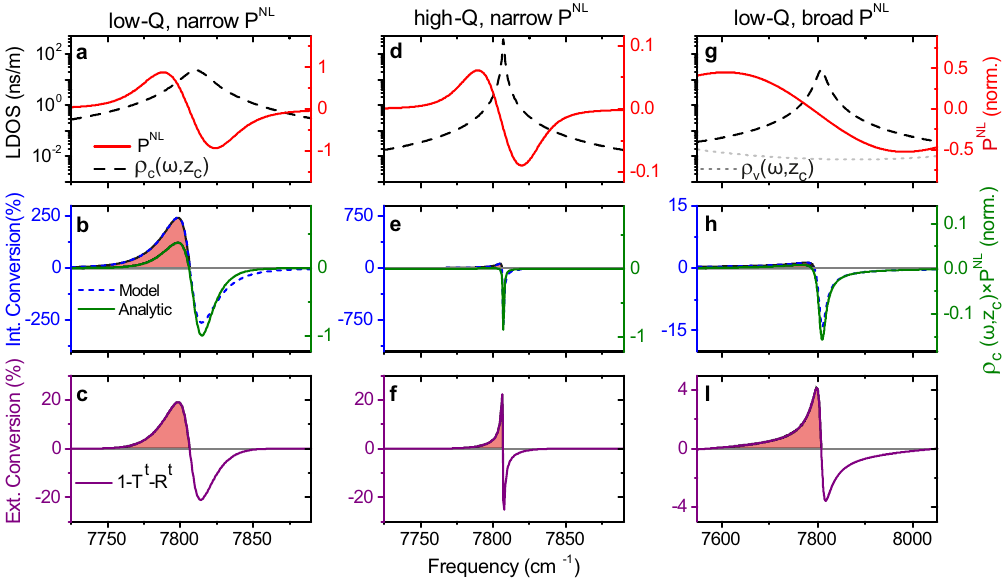}
\caption{\emph{Effects of LDOS and nonlinear polarization on optical frequency conversion. 
Panels (a,d,g) show the LDOS at the cavity center $\rho_c(\omega,z_c)$ (black long-dashed curves), the vacuum LDOS $\rho_v(\omega,z_c)$ (gray dots), the normalized nonlinear polarization ($P^{NL}$) (red curves). 
Panels (b,e,h) show internal color conversion and analytic spectra at the cavity center, where red filled regions highlight the color-converted light. 
Blue short-dash curves are the results of our numerical model, the green curve is the analytic model ($P^{NL} \cdot \rho_c(\omega,z_{c})$) ($P^{NL}$ and $P^{NL} \cdot \rho_c(\omega,z_{c})$ are normalized to the maximum in each row). 
Panels (c,f,i) show color-conversion spectra outside the cavity calculated as $(1 - T^t - R^t)$ (purple curves), where red filled regions highlight the color-converted light. 
In absence of color conversion the external conversion is $(1 - T^t + R^t = 0)$. 
Panels (a-c) pertain to a low-$Q$ cavity ($Q=400$, $ \tau_{c} = \fs{300} $) with a long pump ($\tau_{pu} = \fs{900}$), (d-f) to a high-$Q$ cavity ($Q=7000$, $ \tau_{c} = \fs{6000} $) with the same pump ($\tau_{pu} = \fs{900}$), and (g-i) to the same low-$Q$ cavity ($Q=400$, $ \tau_{c} = \fs{300} $) with a short pump ($\tau_{pu} = \fs{50}$). 
Calculations are performed at $\Delta t = \fs{-30}$ for a probe duration of $\tau_{pr}=\fs{30}$. 
The spectra in (b-i) are normalized to the incident spectrum.}}
\label{DOSCalc}
\end{center}
\end{figure*}

\section{Generalized model}\label{sec:general}
To generalize our new insights, we have performed numerical calculations (see Supplementary Material, Theory). 
We reveal the united effect of both the LDOS and the nonlinear polarization by tuning $Q$ and the switch pulse duration. 
Tuning $Q$ provides the means for changing the bandwidth of the resonant LDOS $\rho_c(\omega,z_c)$ and the amplitude of the vacuum LDOS $\rho_v(\omega,z_c)$.
Independently, we control the bandwidth of the nonlinear polarization $P^{NL}(\omega_{pr})$ in the calculations by tuning the switch pulse duration. 

Results for a low-$Q$ cavity switched by a long pulse are shown in Fig.~\ref{DOSCalc}(a-c). 
We see that the LDOS and the polarization conspire to yield a broad red-shifted frequency conversion down to $\icm{7740}$, with a concomitant blue depletion as far as $\icm{7860}$. 
While one naturally expects a cavity to be active over one linewidth $\Delta \omega_{c}$, Fig.~\ref{DOSCalc} shows that this intuition is erroneous; the bandwidth of frequency conversion greatly exceeds the cavity linewidth ($\Delta \omega_{c} = \icm{18}$), confirming our observation that light is generated in a continuum outside the cavity resonance. 
It appears that a much broader extent of the cavity LDOS spectrum is significant to frequency conversion, which notably includes the vacuum LDOS that tunnels into the cavity $\rho_v(\omega,z_c)$, see Fig.~\ref{DOSCalc}(g). 
These results shed new light on the frequency-conversion spectrum (key feature \# 1): the spectral output is much broader than the resonance and the internal color conversion is given by the product $P^{NL} \cdot \rho_c(\omega,z_c)$, which agrees well with the numerical results.

To elucidate the role of the quality factor $Q$, we show in Fig.~\ref{DOSCalc}(d-f) data for a high-$Q$ cavity with the same nonlinear polarization and mode volume as in Fig.~\ref{DOSCalc}(a-c). 
The LDOS is narrow and strongly pronounced: on resonance the LDOS is an order of magnitude larger, 
and off-resonance an order of magnitude lower 
than in Figure~\ref{DOSCalc}(a), in agreement with the $Q$-dependent LDOS scaling above. 
The LDOS is thus dominated by the cavity resonance, while the role of the vacuum LDOS $\rho_v(\omega,z_c)$ is reduced. 
Indeed, the frequency-converted spectrum in Figure~\ref{DOSCalc}(e) is narrow, mostly within the instantaneous cavity bandwidth, as reported in Ref.~\cite{tanabe.2009.prl}. 
In the time domain, this case means that the cavity storage time has become much longer than the pump pulse duration and hence the switching time $(\tau_{c} >> \tau_{pu})$.  
Spectrally, the LDOS has little overlap with $P^{NL}$, which results in a low frequency-integrated conversion. 
The frequency-integrated conversion Figure~\ref{DOSCalc}(f) is now much smaller than for the low-$Q$ cavity (see Figure~\ref{DOSCalc}(c)), in agreement with the $1/Q$-scaling of our observations in Figure~\ref{DOSMeasCalc}. 
 
To illuminate the role of the nonlinear polarization that is central in traditional nonlinear optics, we present in Figure~\ref{DOSCalc}(g-i) results for a low-$Q$ microcavity (as in Figure~\ref{DOSCalc}(a)), with a broad non-linear polarization $P^{NL}$ (bandwidth $> \icm{400}$). 
In agreement with our analytic model, we find a broader internally generated spectrum (Fig.~\ref{DOSCalc}(h)) and external spectrum (Fig.~\ref{DOSCalc}(i)) compared to Figures~\ref{DOSCalc}(b) and~\ref{DOSCalc}(c).
Due to the fast switching times, the bandwidth of the nonlinear polarization is broader, thus red-shifted light extends beyond $\icm{7600}$, and blue depletion is observed below $\icm{8000}$. 
These results indicate that in the limiting case of a broadband LDOS typical of a homogeneous material, the output spectrum is determined by the nonlinear polarization, which matches the known situation in nonlinear optics in fibers and spatially homogeneous media.

We observe in Figures~\ref{DOSCalc}(b,e,h) a very good agreement between the numerical results and the analytic frequency-conversion proportional to the product of the nonlinear polarization and the LDOS $(P^{NL} \cdot \rho_c(\omega,z_{c}))$. 
Considering in analogy Fermi's golden rule in cQED, one might wonder why the converted intensity is not proportional to the nonlinear polarization squared, assuming $P^{NL}$ to play the role of a transition dipole moment~\cite{gerard.1998.prl, hennessy.2007.nature}. 
For Fermi's golden rule to hold, however, requires the bandwidth of the LDOS resonance to be much broader than the dipole bandwidth, a situation that is known as the weak-coupling or the Markov approximation~\cite{khitrova.2006.natphys}. 
The inequality is strongly violated in the present study, where the polarization bandwidth is comparable to the LDOS bandwidth. 
When the Markov approximation is violated, the intensity may strongly differ from Fermi's golden rule. 
Conversely, for a homogeneous medium with a smooth LDOS, an analogy to Fermi's golden rule should hold; indeed the relations in Ref.~\cite{mizrahi.1988.josab} effectively correspond to a Fermi golden rule. 

\section{Discussion}
\subsection{Unified classification} 

Our new viewpoint yields a natural and unified framework for color-conversion in both traditional and confined media. 
According to our new viewpoint, the frequency-converted signal is the product of the nonlinear polarization change and the LDOS at the new frequencies. 
In practice, as discussed in Figure~\ref{DOSCalc}, the bandwidth of the nonlinear polarization is set by the duration of the switch event (typically the switch pulse). 
The bandwidth of the LDOS is set by the instantaneous resonance frequency of the confined medium (typically a cavity) and its quality factor, or in case of a unconfined medium by the continuum LDOS of free space. 
In case of a single resonance with a relatively high quality factor (see Figs.~\ref{DOSCalc}(d,e,f)), input light is converted to light with a frequency equal to the instantaneous cavity resonance, since the LDOS outside the cavity resonance is low; there are no other frequencies available to the converted light than the cavity resonance. 
Interestingly, this situation corresponds to the cases in Refs.~\citep{notomi.2006.pra, tanabe.2009.prl} that were referred to as "adiabatic". 

Let us now discuss the situations in Figs.~\ref{DOSCalc}(d,e,f) versus Figs.~\ref{DOSCalc}(g,h,i): in both cases the switch event is faster - by about $6 \times$ - than the cavity storage time $(\tau_{pu} < \tau_{c})$. 
Nevertheless, remarkable differences appear: 
in Figs.~\ref{DOSCalc}(d,e,f), the frequency converted spectrum is relatively narrow on account of a narrow LDOS bandwidth (due to the elevated quality factor $Q$). 
In contrast, in Figs.~\ref{DOSCalc}(g,h,i) the frequency converted spectrum is relatively broad on account of a broad LDOS bandwidth (due to the low quality factor $Q$). 
These differing results are at variance with previous work~\cite{notomi.2006.pra} that claimed that so-called non-adiabatic frequency conversion (\textit{i.e.}, conversion outside the resonance) occurs in the limit $\tau_{pu} < \tau_{c}$. 
This contradiction can only be resolved by concluding that the hypothesis of Ref.~\cite{notomi.2006.pra} is invalid. 
Based on our experiments as a function of quality factor and our theoretical findings, we see that in these two cases, where $\tau_{pu} < \tau_{c}$, different frequency conversion occurs: either close to the cavity resonance ("adiabatic") or far from the resonance ("non-adiabatic"), and the outcomes are in this limit (where the nonlinear polarization spectrum is broad) determined by the bandwidth of LDOS. 
 
When a cavity has multiple resonances, the LDOS has substantial contributions in between the resonances, on account of neighboring resonances. 
Hence, light is readily converted to frequencies in between the resonances, a situation that corresponds to the non-adiabatic situation presented in Ref.~\cite{dong.2008.prl}. 

\subsection{Applications} 
The unified framework introduced here describes nonlinear frequency-conversion in open and confined systems. 
Our theory not only explains our experimental results but also elucidates earlier results in the literature. 
Recently, we observed that the electronic Kerr effect allows for the repeated switching of a cavity resonance at THz rates~\cite{yuce.2013.ol}. 
Therefore, we project that trains of blue- and red-shifted pulses can be generated at THz repetition rates. 
The significance of the environment can also serve to explain the complex nature of supercontinuum generation by implementing models that allow to calculate the LDOS in 3D~\cite{nikolaev.2009.josab}. 
Moreover, one could even \textbf{design} converted light to appear in targeted bands, and possibly even with pre-designed spectral shapes, desired for frequency combs \cite{Yu.2016.Optica, Stern.2018.Nature} using control over the LDOS and switch pulse shape. 
Here, we enter a new dimension in the control of the color of light - beyond the well-known control means~\cite{boyd.2008.book} - namely by controlling and engineering the LDOS of the nonlinear medium.
Since in case of Kerr switching the probe light stored in the cavity only experiences a real refractive index, the prospects are favorable to employ cavity-aided frequency conversion to single photons and other quantum states of light~\cite{englund.2012.prl, volz.2012.nat.phot.}, thus paving the road to novel opportunities in quantum optics and cQED.

\section*{Funding Information}
This research was supported by Smartmix-Memphis, NWO-FOM-projectruimte (“Zap!”: Ultrafast time control of spontaneous emission), ERC ("Pharos", PI Mosk), NWO-Nano ("Reversible Slowing of Light by Nanophotonic Phase Imprinting", PI Mosk, co-PI WLV), and NWO-STW ("Dispersion engineering of random photonic media for improved white light emitting diodes"), as well as Philips Lighting (Signify), Lumileds, Thales France, and METU-BAP Project: 2665. 

\subsection*{Acknowledgments}
It is a pleasure to thank Alex Gaeta, Ad Lagendijk, Allard Mosk, Emanuel Peinke, and Henri Thyrrestrup for insightful discussions. 
An early version of this manuscript was posted on the Cornell preprint server (arXiv:1406.3586).

\section*{Supplemental Documents}
See Supplement 1 for supporting content.

\bibliography{C:/Users/eyuce/Dropbox/Library/151115_references}

\end{document}